\documentclass[12pt]{article}
\usepackage[utf8]{inputenc}
\usepackage{graphicx}
\usepackage{amsmath}
\usepackage{hyperref}
\usepackage{geometry}
\usepackage{authblk}
\usepackage{float}
\usepackage{booktabs}
\title{Using text embedding models as text classifiers with medical data}

\author[1]{Rishabh Goel}
\author[2]{Ramin Ramezani}
\affil[1]{Monta Vista High School, Cupertino, California}
\affil[2]{University of California Los Angeles, Los Angeles, California}

\date{July 7, 2024}

\begin{document}

\maketitle

\begin{abstract}
The advent of Large Language Models (LLMs) is promising and LLMs have been applied to numerous fields. However, it is not trivial to implement LLMs in the medical field, due to the high standards for precision and accuracy. Currently, the diagnosis of medical ailments must be done by hand, as it is costly to build a sufficiently broad LLM that can diagnose a wide range of diseases. Here, we explore the use of vector databases and embedding models as a means of encoding and classifying text with medical text data without the need to train a new model altogether. We used various LLMs to generate the medical data, then encoded the data with a text embedding model and stored it in a vector database. We hypothesized that higher embedding dimensions coupled with descriptive data in the vector database would lead to better classifications and designed a robustness test to test our hypothesis. By using vector databases and text embedding models to classify a clinician's notes on a patient presenting with a certain ailment, we showed that these tools can be successful at classifying medical text data. We found that a higher embedding dimension did indeed yield better results, however, querying with simple data in the database was optimal for performance. We have shown in this study the applicability of text embedding models and vector databases on a small scale, and our work lays the groundwork for applying these tools on a larger scale.
\end{abstract}

\section{Introduction}
The rapid advancement of medical knowledge creates a need for new ways to store and represent medical data. For example, one rapidly advancing area is natural language processing (NLP) and the advancements have provided invaluable tools, namely large language models (LLMs), which are built off the transformer architecture and can encode and process non-numerical data in a numerical way \cite{vaswani2017attention}.

However, LLMs are not the only major advancement in NLP in recent times. Vector databases and text embedding models have advanced in recent years and are increasingly usable in various contexts, such as medicine \cite{stata2000term, collobert2008unified}. For example, researchers at Google have developed an LLM, Med-PaLM, to explore the uses of LLMs specifically in the medical context \cite{singhal2023clinical, chowdhery2023palm}. Text embedding models and vector databases provide a robust method of both storing and representing non-numerical data, such as text data. Utilization of these tools in areas that use a large amount of text, like clinical data, should be tested and their robustness quantified, given their potential use in medical circumstances.

Recent medical tools, such as Med-PaLM, rely on large
amounts of training data and validation to be successful
in communicating with clinicians and providing accurate
diagnoses and information for general applications \cite{chowdhery2023palm}.
Otherwise, AI in the field of medicine has been limited to
niche use cases, where models are only trained for a specific
task within the field of pathology, such as analyzing lymph
node biopsies for metastatic breast cancer \cite{ahmad2021artificial}. Furthermore,
there is a lack of annotated data for model training \cite{ahmad2021artificial}.

This study focused on using the existing abilities of text
embedding models to accurately classify textual data in
a medical context, without the need to train specifically for
medical text data or rely on annotated data. The hope was
that through testing with different kinds of text data, we could
reach a qualitative and quantitative conclusion about the
type of text data that can best achieve accurate diagnoses.
However, since this data requires collection by medical
professionals, it was difficult to obtain. Therefore, we used
the existing knowledge embedded in LLMs and used that
to test the applicability of text embedding models as text
classifiers in medicine. This would allow clinicians to classify
their text and get diagnosis help without the need to interact
with an LLM. Additionally, by examining what type of text data
(whether sparse or detailed), we also reached conclusions on
the optimal way to store and retrieve data to make the best
classifications. Since greater detail and more context usually
leads to better sentence prediction for LLMs, we felt that it
was worth exploring if the same behavior translated to text
embedding models and vector databases \cite{petroni2020context}.

We designed a robustness test to measure the quantitative
aspects of the textual data presented. We hypothesized that
higher embedding dimensions, coupled with descriptive data
in the vector database, would lead to better classifications of
medical data. Through our experimentation, we showed that
having a large amount of data stored in the vector database
was far more effective for classifying text than having a
less data. More data in the vector database also allowed
for sparser data to query it, as the vector database’s cosine
similarity was better able to accurately classify a sparse query
when comparing it to a dense vector.

\section{Materials and Methods}

\begin{figure}[H]
    \centering
    \includegraphics[width=\textwidth]{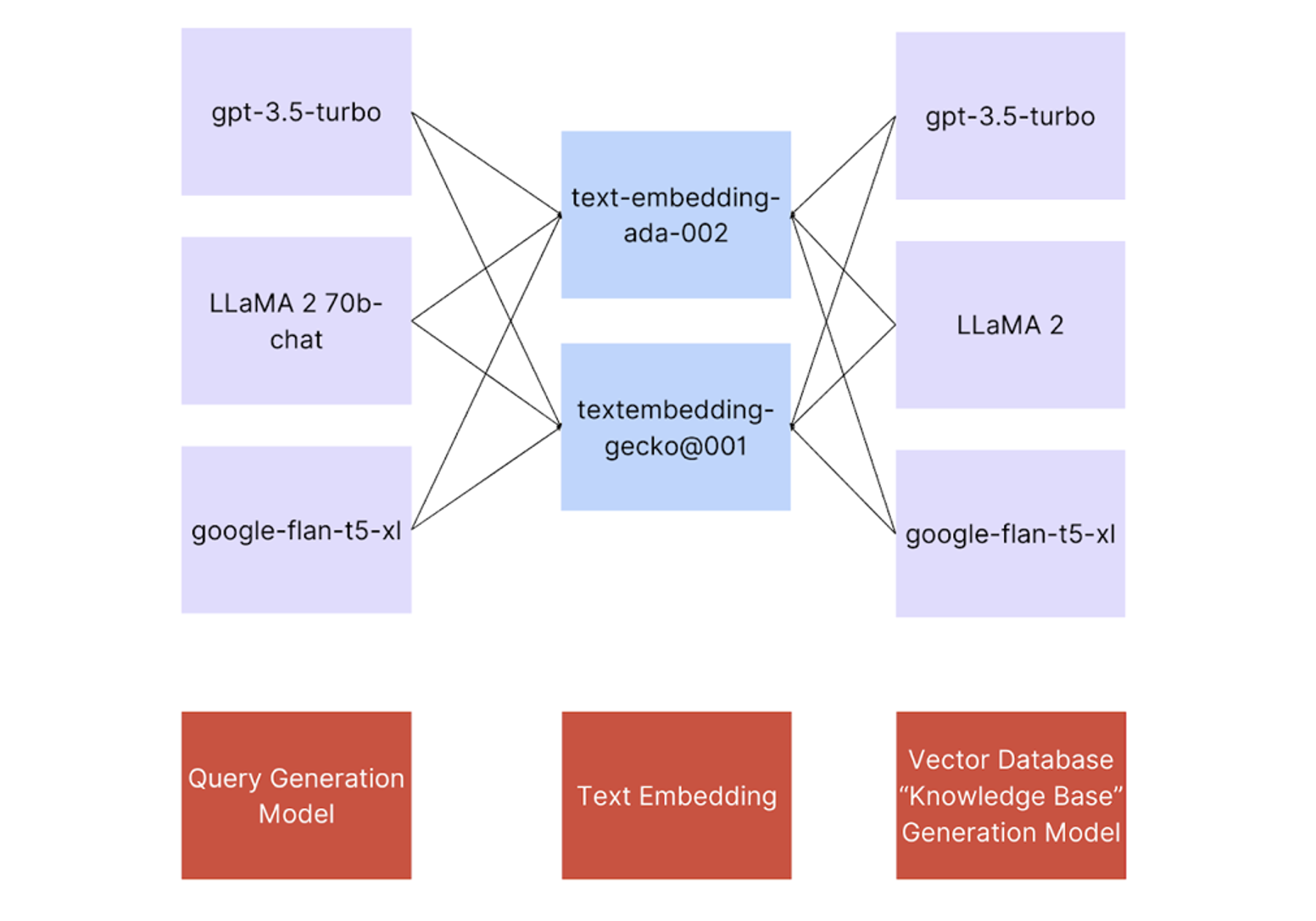}
    \caption{Flow chart of robustness test pipeline. There are 3x2x3 = 18 permutations of tests possible. Each query generation model was paired up against each text embedding model and each knowledge generation model. Embeddings from the query generation model were compared with the embeddings from the knowledge generation model using a cosine similarity.}
    \label{fig:flowchart}
\end{figure}

\subsection{Vector Database}
We began by treating the vector database as a knowledge base. Here, the vector database for each ailment contained correct medical knowledge and was prompted to include observable symptoms of a particular ailment, as well as a potential sample or ranges of tests and expected abnormalities. The classification tasks we were trying to solve were the abilities of vector databases and vector embedding models to act as adequate text classifiers. To accommodate this, we needed a labeled dataset that was correct for the vector database and the embedding model to use as the standard. For this reason, the vector database stored the vectors that were supposed to be the ground truths, which corresponded to the correct medical classifications. To query and get results, we had a list of notes or test results that are commonly produced during examination by a clinician. These were then compared with the ground truth values in the vector database and the vector database would return the vector that the similarity search had evaluated to be the most similar. Since the similarity scores and the vectors that were returned from the vector database did not contain information on which ailment it corresponded to, we added metadata to each vector in the database. The metadata that was tied to this vector was used to determine which ailment has been selected and the classification decision of the vector database.

\subsection{Workflow}
As a general framework, data was generated using a prompt on an LLM, which was then represented as a string. We then took these results from the LLMs and fed them into the text-embedding model, where the string data was converted into a vector representation of the text. We did this multiple times for each ailment (Table \ref{tab:dataset}). Then, one vector for each ailment was stored in the Pinecone vector database as the ground truths or the knowledge base. Once there were ground truths in place, we queried the remaining vectors on those in the vector database. The query returned the vector in the database that the query vector was most similar to and a similarity score, which quantified their similarity. If the most similar vector and the query vector were of the same ailment (with a similarity score of at least 0.5), this was counted as a positive classification.

We used three models for data generation: OpenAI's gpt-3.5-turbo, Meta's LLaMA 2 70b-chat, and Google's flan-t5-xl. To generate the data, we used OpenAI's API for gpt-3.5-turbo outputs and HuggingFace for LLaMA 2 70b-chat and google-flan-t5-xl \cite{chung2022scaling,touvron2023llama,brown2020language}. To classify the text, we used two text embedding models: OpenAI's text-embedding-ada-002 and Google's textembedding-gecko@001. Again, for the OpenAI model we used the API and for textmebedding-gecko@001 we used Google Cloud SDK. All code was written in Python, and we used Seaborn and Matplotlib to generate the confusion matrices.

\subsection{Data}
We conducted the test on eight ailments: glaucoma, jaundice, cyanosis, psoriasis, conjunctivitis, scoliosis, skin cancer, and gingivitis (Table \ref{tab:dataset}). Each ailment had one vector in the vector database. Since the vector in the vector database was acting as the ground truth of medical knowledge, there did not need to be many vectors and instead each vector was generated with text that was prompted to be comprehensive and with a maximum token size of 500 where a token was roughly three to four characters. The number of queries per ailment was far higher but with shorter text at only a maximum of 50 tokens per query. A large amount of data allowed for a large amount of testing for each of the ailments. However, the time to generate queries was quite long, so we limited the maximum number of queries per ailment. The individual queries were designed to have a low amount of data in them to really test the capabilities of the embedding model and the vector database and their ability to classify the vector in a fashion that preserved meaning.

When querying the vector database and getting results using the textembedding-gecko@001 model, due to API request limits, we had to truncate the number of queries that were generated per LLM. We ended up using one third of the queries that were generated by both gpt-3.5-turbo and LLaMA 2 70b chat and one fifth of the queries that were generated by google flan-t5-xl since it had more queries.

\subsection{Prompt Engineering}
We treated the prompts as constant, as we did not want much variance in how the model structured its response, but we wanted variance in the data that was generated. Since the vector database acted as the ground truth for medical knowledge, we made this part of the prompt the most comprehensive and inclusive. The prompt for generating the ground truths for LLaMA 2 70b-chat and gpt-3.5-turbo that were then placed into the vector databases was as follows:

``What notes would a doctor have when observing a patient with a \{particular ailment\}? Include test results. Do not include the patient's name, age, gender, or any patient specific details including date of observation. Do not tell me that the notes are not comprehensive, I already know this. Do not tell me about anything that requires further investigation.''

The prompt for generating each of the queries was slightly different. Here, an emphasis was placed on being vague to mimic the reality of patient observation. The generation prompt was as follows:

``What notes would a doctor have when observing a patient with \{a particular ailment\}? Make notes short and concise. Laboratory test results are optional. Do not include the patient's name, age, gender, or any patient specific details including date of observation. Do not tell me that the notes are not comprehensive, I already know this. Do not tell me about anything that requires further investigation. Do not tell me what ailment the patient is presented with.''

The prompt for the query generation was remarkably similar to the prompt for generating ground truths, except that it made test results optional and put an emphasis on short and concise notes.

Since google-flan-t5-xl outputs extremely short answers, the prompt was modified:

``What are observable symptoms of \{particular ailment\}? List them out to be technical.''

There was no mention of laboratory results because the model would not generate them. This sharp contrast in data format provided a good test for the embedding model.

The results from the gpt-3.5-turbo and LLaMA 2 70b-chat query prompts were long lists. To reduce the uniformity of data and only give the text embedding model a small amount of context, the lists that were generated were divided so that each query consisted of three items from the list. This reduced the scope that the embedding model had when comparing the query to the ground truth and allowed for more queries with fewer API calls. The splitting process was as simple as just taking every three items in order and embedding those.

\subsection{Robustness Test Definition}
To address the issue of the dataset being generated by LLMs instead of medical professionals, we present a method of determining the efficacy of text embedding models and the vector databases that is not as heavily reliant on the quality of the data presented. The text embedding model classified the vector data by determining whether the two instances of text were similar. If the data carried similar meaning, then they were embedded close to each other or the angle between the vectors was small. If the embedding model could handle inputs from different LLMs, the outputs of which consisted of varying levels of length and specificity, then the embedding model should have been able to find similarities in data even though the ideas presented were represented in different ways. By pairing up each possible query generating LLM to each possible ground truth generating LLM and using each text embedding model to classify the text generated by both models, we were able to see how each text embedding model handled each LLM's nuances in generating data, and how well it was able to embed the same data represented in different ways. Each one of these permutations of an LLM/text embedding model combination was a robustness test. A total of 18 different permutations of robustness tests were possible, involving combining three LLMs (gpt-3.5-turbo, google-flan-t5-xl, LLaMA 2 70b-chat) twice (once for query generation and once for ground truth generation) and two text embedding models (text-embedding-ada-002, textembedding-gecko@001) (Figure \ref{fig:flowchart}). Each robustness test has been represented as a confusion matrix with dimensions of 8x8 for the eight ailments.


\begin{table}[h]
\centering
\caption{Dataset overview. Number of queries per LLM per ailment}
\label{tab:dataset}
\resizebox{\textwidth}{!}{%
\begin{tabular}{@{}lcccccccc@{}}
\toprule
 & Glaucoma & Jaundice & Cyanosis & Psoriasis & Conjunctivitis & Scoliosis & Skin Cancer & Gingivitis \\
\midrule
gpt-3.5 queries & 135 & 100 & 115 & 127 & 136 & 127 & 145 & 117 \\
flan-t5 queries & 200 & 200 & 200 & 200 & 200 & 200 & 200 & 200 \\
LLaMA queries & 121 & 157 & 124 & 129 & 135 & 116 & 100 & 130 \\
\midrule
Total & 456 & 457 & 439 & 456 & 471 & 443 & 445 & 447 \\
\bottomrule
\end{tabular}%
}
\end{table}

\section{Results}

We performed this study in hopes of understanding how embedding dimensions along with density of knowledge in our knowledge base would influence the quality of classification by the text embedding models. We quantitatively validated our hypothesis using a robustness test, which provided us with standard classification metrics --- such as accuracy and F1 score --- so that we could compare the performance of different models. Instead of training a multi-billion parameter LLM locally, we explored the effectiveness of existing tools --- namely vector databases and text embedding models --- in the medical context \cite{hoffmann2022empirical}. The LLMs used to generate the data were OpenAI's gpt-3.5-turbo, Google's flan-t5-xl, and Meta's LLaMA-70b-chat \cite{chung2022scaling,touvron2023llama,brown2020language}. The text embedding models used to classify the data were OpenAI's text-embedding-ada-002 and Google's textembedding-gecko@001. There are 18 possible robustness tests, with three LLMs to generate queries and ground truths and two text embedding models, all of which may be grouped independent of each other (3 x 3 x 2 = 18). For brevity's sake, we did not analyze all robustness tests, and only chose the ones we believed were the most interesting to focus on (Appendix). We also only focused on eight ailments in this study: glaucoma, jaundice, cyanosis, psoriasis, conjunctivitis, scoliosis, skin cancer, and gingivitis which were chosen due to their commonality, where it was likely that the LLMs would have knowledge of them, as well as partial overlap of their symptoms. For example, skin cancer and psoriasis can have similar symptoms, and we were interested to see how the LLMs would perform in such contexts.

We used an LLM to generate a query dataset and a ground truth dataset, which we then stored in a vector database. Next, we embedded each of the queries using a text embedding model in the query dataset. We then compared each query vector to the vectors in the vector database, and the most similar one was chosen. If the chosen vector was describing the same ailment as the query vector, then we counted this as a positive classification (Figure \ref{fig:flowchart}).

\subsection{High detail in both knowledge and query dataset}

\begin{figure}[H]
    \centering
    \includegraphics[width=\textwidth]{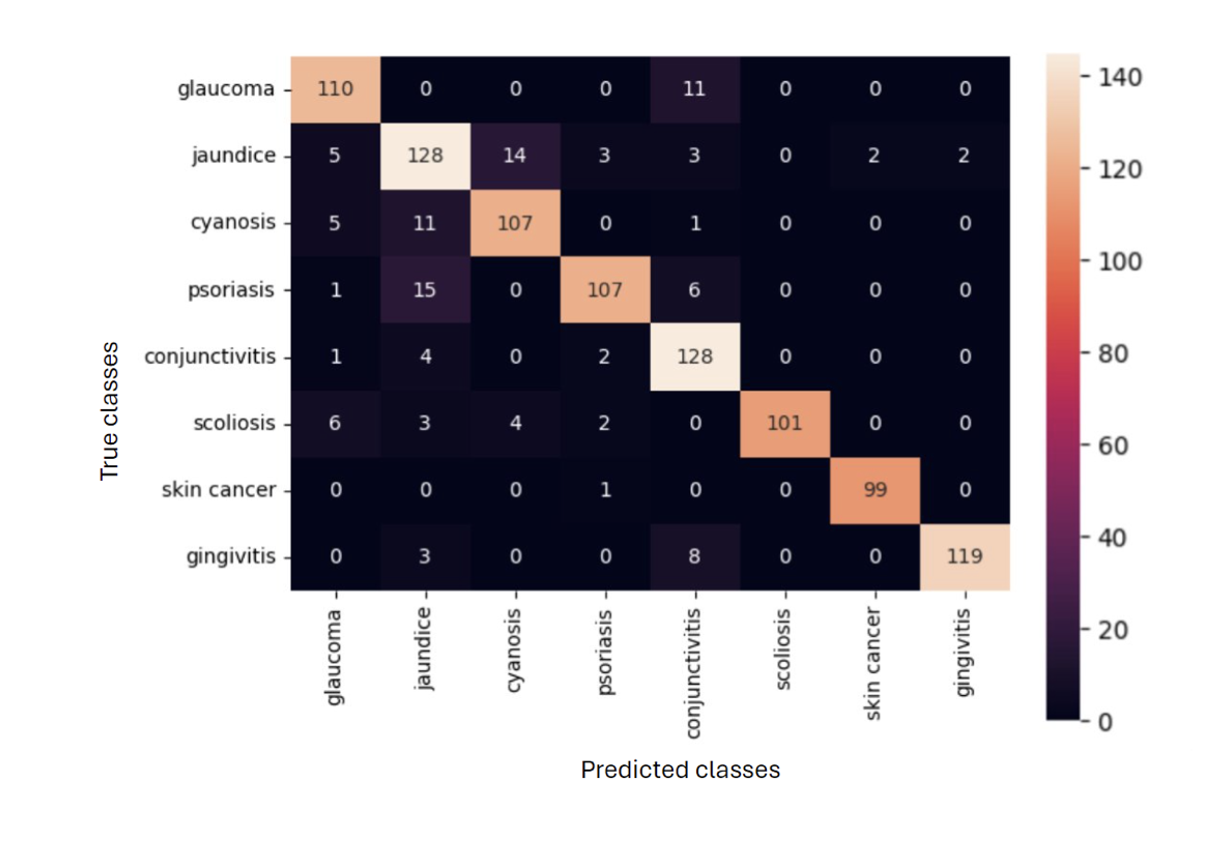}
    \caption{Confusion matrix describing the results from the first robustness test where there were detailed querying and knowledge bases. Rows describe the true values, and columns describe the predicted values. The query dataset was generated by LLaMA 2 70b-chat, text-embedding-ada-002 was used as the embedding model, and the ground truth dataset in the vector database was generated by gpt-3.5-turbo.}
    \label{fig:matrix1}
\end{figure}

First, we examined a situation where two extremely high detail LLMs generated both query and ground truth databases (Figure \ref{fig:matrix1}). In conjunction with this, the vector embedding model also preserved the most information as its dimensionality was the highest (1536) compared to textembedding-gecko@001, which has a dimensionality of 768. Consequently, the classification metrics were positive with a misclassification rate of only 11\% and a macro F1 score of 0.89.

\subsection{Using sparse querying with detailed knowledge base}

\begin{figure}[H]
    \centering
    \includegraphics[width=\textwidth]{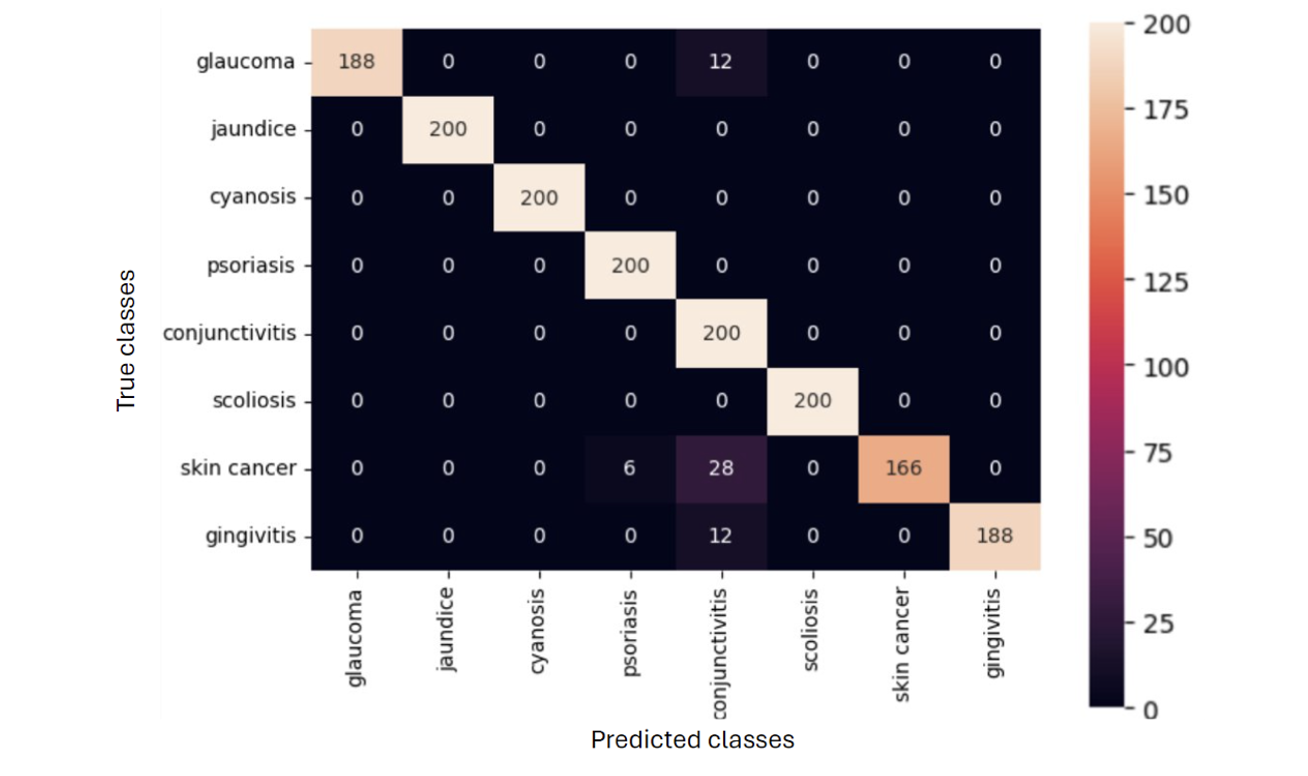}
    \caption{Confusion matrix describing the results from the second robustness test where there was a sparse querying base and a detailed knowledge base. Rows describe the true values, and columns describe the predicted values. The query dataset was generated by google-flan-t5-xl, text-embedding-ada-002 was used as the embedding model, and the ground truth dataset in the vector database was generated by gpt-3.5-turbo.}
    \label{fig:matrix2}
\end{figure}

Next, we examined a situation where a less detailed LLM, flan-t5-xl, created the query set, and the ground truth set was still created by a very high detail LLM, gpt-3.5-turbo (Figure \ref{fig:matrix2}). Again, the embedding model used was text-embedding-ada-002, which has high dimensionality. The combination of these models resulted in a low misclassification rate of only 3.63\% and a macro F1 score of 0.96, far outperforming those situations where gpt-3.5-turbo generated the ground truth dataset and LLaMA-70b-chat generated the query dataset (Figure \ref{fig:matrix1}).

\subsection{Sparse data as knowledge base}

\begin{figure}[H]
    \centering
    \includegraphics[width=\textwidth]{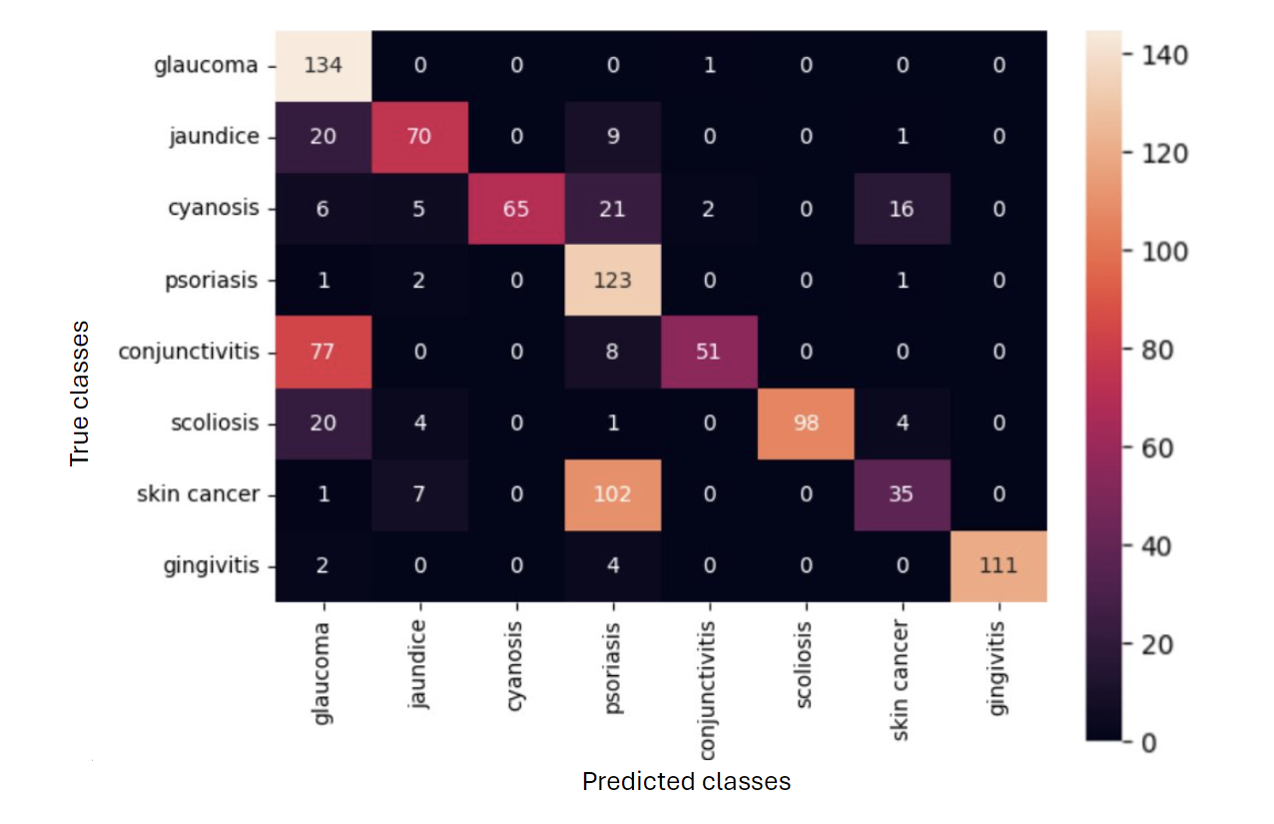}
    \caption{Confusion matrix describing the results from the third robustness test where there was a detailed querying base and a sparse knowledge base. Rows describe the true values, and columns describe the predicted values. The query dataset was generated by gpt-3.5-turbo, text-embedding-ada-002 was used as the embedding model, and the ground truth dataset in the vector database was generated by google-flan-t5-xl.}
    \label{fig:matrix3}
\end{figure}

Next, gpt-3.5-turbo (a detailed LLM) created the query dataset, which was paired with the ground truth database that was generated by flan-t5-xl, which usually generates sparse or vague responses (Figure \ref{fig:matrix3}). Despite text-embedding-ada-002 being used, the high dimensionality was not enough to clearly distinguish between ailments as the misclassification rate was 31.4\%. 25\% of ailments had more misclassifications than actual classifications. The ailments that did not have high misclassification numbers --- namely psoriasis and glaucoma --- had many false positives except for gingivitis, with glaucoma having a precision of 0.513 and psoriasis having a precision of 0.459.

\subsection{Same model for knowledge and query dataset}

\begin{figure}[H]
    \centering
    \includegraphics[width=\textwidth]{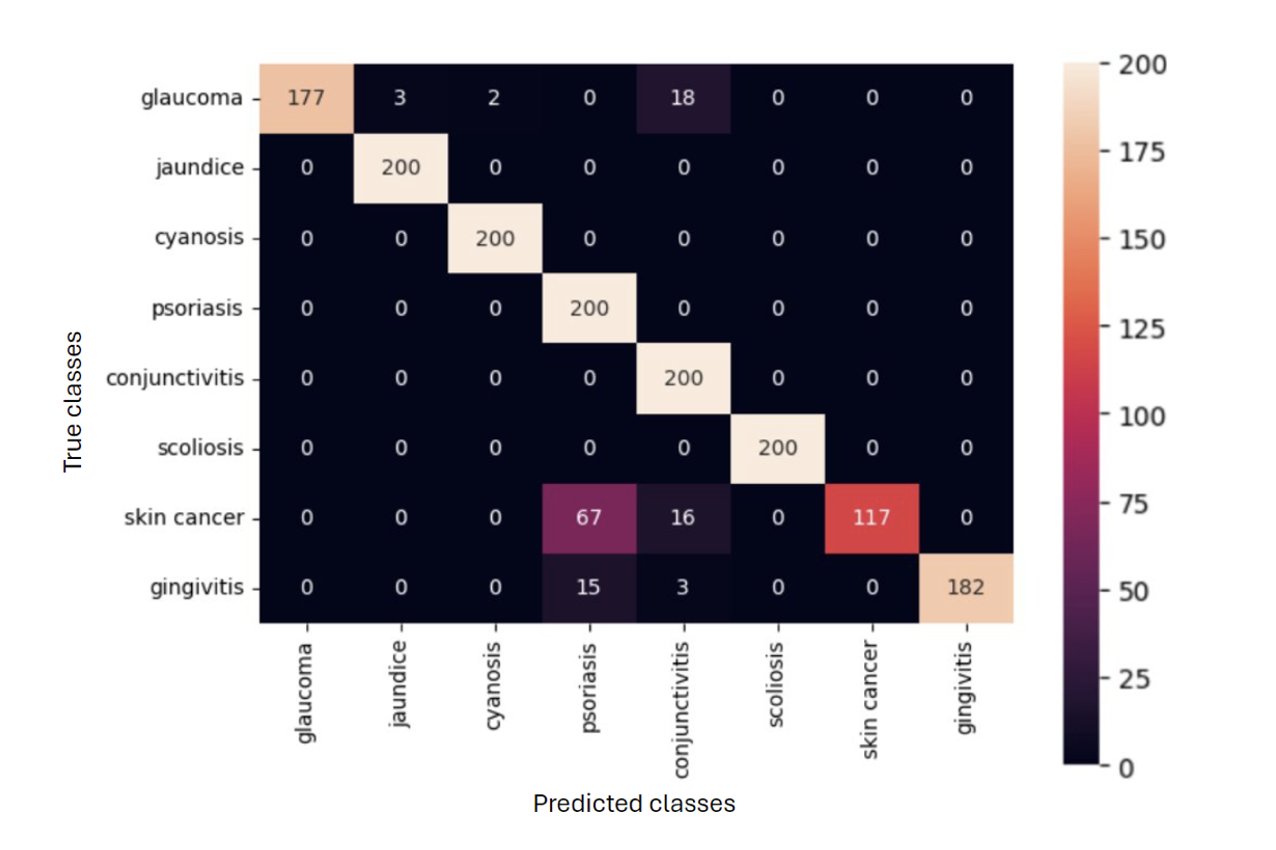}
    \caption{Confusion matrix describing the results from the fourth robustness test where there were sparse querying and knowledge bases. Rows describe the true values, and columns describe the predicted values. The query dataset was generated by google-flan-t5-xl, text-embedding-ada-002 was used as the embedding model, and the ground truth dataset in the vector database was generated by google-flan-t5-xl.}
    \label{fig:matrix4}
\end{figure}

We then performed a test where flan-t5-xl generated both the query dataset and the ground truths (Figure \ref{fig:matrix4}). Despite the ground truths and queries being generated by the same LLM, there still were some misclassifications, especially with skin cancer as it had a recall of only 0.585. Furthermore, psoriasis had quite a few false positives, resulting in a low precision of 0.701. Apart from these two notes, the classification results were overall quite strong with a macro F1 score of 0.92. Interestingly, the flan-t5-xl model performed worse when it was querying on vectors generated by itself than when it was querying on vectors generated by gpt-3.5-turbo.

\subsection{Effects of vector dimensionality}


\begin{figure}[H]
    \centering
    \includegraphics[width=\textwidth]{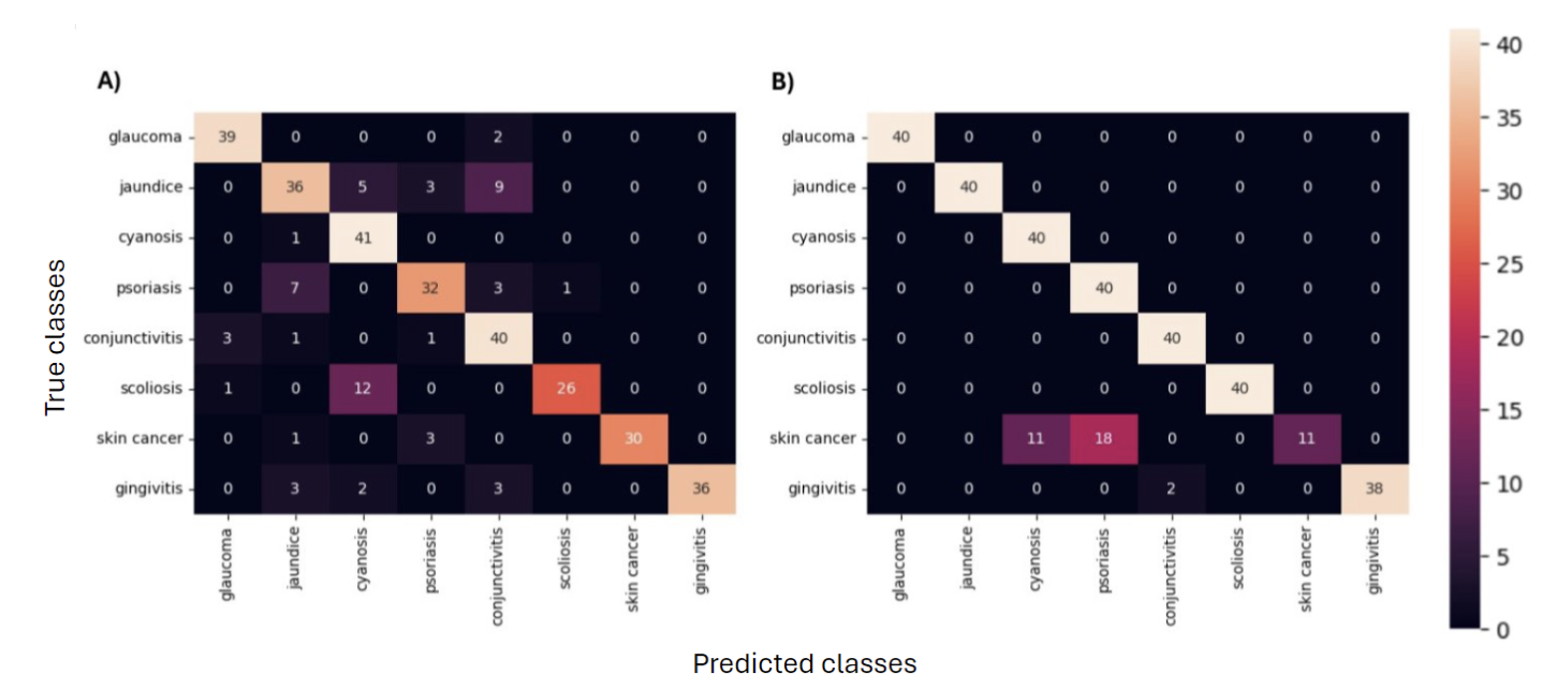}
    \caption{Confusion matrix describing the results from the fifth and sixth robustness tests. Rows describe the true values, and columns describe the predicted values. (A) The query dataset was generated by LLaMA 2 70b-chat, textembedding-gecko@001 was used as the embedding model, and the ground truth dataset in the vector database was generated by gpt-3.5-turbo. (B) The query dataset was generated by google-flan-t5-xl, textembedding-gecko@001 was used as the embedding model, and the ground truth dataset in the vector database was generated by gpt-3.5-turbo.}
    \label{fig:matrix5&6}
\end{figure}

To understand the effects of vector dimensionality on classification, we repeated the first two tests using textembedding-gecko@001 rather than text-embedding-ada-002 (Figure \ref{fig:matrix5&6}). The textembedding-gecko@001 embedding model has a dimensionality which is half that of text-embedding-ada-002. Even with two detailed LLMs, here, the misclassification rate was 17.9\% in comparison to the 11\% with text-embedding-ada-002 (Figure \ref{fig:matrix5&6}a). The recall for skin cancer when textembedding-gecko@001 was used was 0.88 as compared to 0.99 when text-embedding-ada-002 was used (Figure \ref{fig:matrix5&6}b, Figure \ref{fig:matrix2}).

\section{Discussion}

Considering the large costs of training a medical LLM to quantify the validity of medical data, we used existing text embedding models and medical data to create text classifiers and tested their performance. We hypothesized that higher embedding dimensions, coupled with descriptive data in the vector database, would lead to better classifications of medical data. We tested this with a robustness test to measure both the qualitative and quantitative aspects of the textual data presented. In medical fields, the accuracy of information is paramount. Along with accuracy of the information, it is important that the information presented is fair and without bias \cite{omiye2023large}. We designed the robustness test with this in mind. We showed that descriptive data stored in the vector database is more effective for classifying text than sparse data. Here, we discuss the implications of each test and the specific behaviors of each LLM and text embedding model.

We acknowledge that our method of generating medical ground truths through LLMs is unconventional and may be slightly inconsistent with accepted medical knowledge, which does not make the use of LLMs necessary for a replication of this study. In fact, using a dataset curated by medical professionals would have been ideal as it would provide more realistic test cases. However, with the emergence of LLMs that specialize in the medical field that can act as a source of medical knowledge, such as Google's Med-PaLM, such LLMs can generate data that agrees with accepted medical knowledge \cite{singhal2023clinical}. Due to constraints relating to a lack of availability of data that provide accurate test results and comprehensive observable symptom lists for specific ailments, we opted for the generation of this data rather than using a curated dataset. This generation of data also allowed for another model component for us to vary, which was the LLM being used to generate the data itself.

It is not surprising if a text embedding model performs better, or has a higher accuracy and F1 score, while classifying text generated from the same model. What is more interesting is how these models performed when queried with data from other models. Even though all these models are LLMs, they still varied greatly in their representation of the same data, especially in the case of google-flan-t5-xl as compared to gpt-3.5-turbo and LLaMA 2 70b-chat.

The google-flan-t5-xl model unsurprisingly performed quite well when it was queried with data generated by itself. However, despite the content, length, and level of detail being nearly identical in the query and ground truth datasets, it still performed poorly in classifying ``skin cancer''. The google-flan-t5-xl model misclassified 83 queries and only classified 117 skin cancer cases correctly, mostly misclassifying skin cancer as psoriasis, likely due to the model generating similar text for both the ailments. This, however, is not very representative of the real world as it is unlikely that the medical ground truth dataset will be as sparse as the data that is used to query it.

To assess performance when the ground truth and query databases were generated with different models, we began testing by keeping the ground truth database generated by the gpt-3.5-turbo model. This model is known for being a conversational LLM, and therefore has quite lengthy responses. As seen later, having a comprehensive and elaborate ground truth database made for a good ground truth model. The LLaMA 2 70b-chat model is like gpt-3.5-turbo, in the sense that it is also a conversational LLM and, therefore, exhibits the same types of characteristics. It, however, is not as powerful and has a smaller context window so it does not go into as much detail as gpt-3.5-turbo does, although it is much closer in detail to gpt-3.5-turbo than to google flan-t5-xl. The LLaMA 2 queries (which, as mentioned above, were only three listed symptoms/lab results) performed quite well with the gpt-3.5-turbo generated ground truth database. While these two models' responses had comparable length, they provided different information and LLaMA 2 was not as detailed, with LLaMA 2 mentioning certain symptoms or phrasing them in such a way that gpt-3.5-turbo would not. Even with this discrepancy in style and detail, the text-embedding-ada-002 model was able to sufficiently classify texts from these two models, with a low misclassification rate of 11\%.

What is more surprising is the efficacy of the google-flan-t5-xl model at generating queries when gpt-3.5-turbo has generated the ground truths. The google-flan-t5-xl model generated exceptionally short queries, with each query being two words to a sentence long. The text-embedding-ada-002 model performed exceptionally well when it came to classifying the queries generated by the google-flan-t5-xl model, with only 58 misclassifications out of 1600 queries. The text-embedding-ada-002 model and the vector database seemed to have an easier time finding similarities when the data in the vector database was more elaborate and comprehensive and the querying data was much shorter and more concise. This is good news because the ground truth of medical data is going to be comprehensive and elaborate, whereas a clinician's notes about a patient exhibiting a certain ailment will typically not be as detailed. Despite the highly distinctive styles of information that was being presented in the data generated between google-flan-t5-xl and gpt-3.5-turbo, the embedding model did well at finding similarities.

It is important to note, however, that because the google-flan-t5-xl model would output such short messages, it often output just the key words associated with a particular ailment. For example, for scoliosis, it output ``bending spine.'' This itself is a keyword and made it slightly easier to classify. This is a possible explanation for why the google-flan-t5-xl model performed better than the LLaMA 2 70b-chat model, even though the latter is significantly more powerful. Another possible explanation is that the google flan-t5-xl was prone to repeating these keywords, despite the temperature being set quite high (we set it at 1.5). The LLaMA 2 70b-chat model did not give a more definitive answer, such as ``bending spine'' for scoliosis, but rather had a more speculative approach when observing characteristics, which may be more realistic. In a real medical setting, the performance of the google-flan-t5-xl model would not be as excellent as it was here. Nonetheless, it is impressive that the text-embedding-ada-002 model and the Pinecone vector database could correctly match these messages that differed in length and detail, with a misclassification rate of 3.63\%.

The results changed when google-flan-t5-xl became the model that generated the data for the ground truths. This data was so sparse and lacking in detail that many of the queries generated by the other, more descriptive models could fit for several ailments that the google-flan-t5-xl had generated as ground truth. This can be seen quite clearly when looking at ``skin cancer''. The text-embedding-ada-002 model and the vector database only correctly classified 35 cases of skin cancer and misclassified 110, with 102 of those being misclassified as psoriasis, which is an ailment that also affects the skin. The queries generated by the gpt-3.5-turbo and the LLaMA 2 70b-chat models also had laboratory test results that would correspond with a certain ailment, whereas the ground truth dataset that was in the vector database did not have any lab test results to compare it to. Therefore, all queries that had only lab test results in them were most likely just guesses.

Overall, the google-flan-t5-xl model is not robust enough to generate a medical ground truth dataset. The embedding model and the vector database performed far better and were more easily able to distinguish between ailments, especially similar ones, when the data in the vector database was far more descriptive. However, it is quite unlikely that the ground truth of medical knowledge will be short and lacking in detail, as there is an extensive supply of detail that is ever growing. Thus, google-flan-t5-xl is neither an accurate representation of real medical knowledge nor is it effective enough to act as one.

Another large part of our experiment was discussing the role text embedding dimensionality has in performance. All the results discussed above were embedded using OpenAI's text-embedding-ada-002 model. This model has a dimensionality of 1536, twice that of Google's Vertex AI's textembedding-gecko@001, which has a dimensionality of 768. This doubling of dimensionality resulted in a slight, but not unimportant, improvement in the ability of the vector database to find similarities between two vectors. This finding makes sense because the text-embedding-ada-002 model could encode more information into its vectors, and therefore it was able to account for more nuanced details than textembedding-gecko@001 could.

We initially hypothesized that a higher embedding dimension coupled with descriptive data, both in the knowledge and query datasets, would yield better results. Indeed, we saw that the text-embedding-ada-002 model with 1536 embedding dimensions performed far better than textembedding-gecko@001 on many occasions. However, the data did not need to be very descriptive when it was used to query the knowledge base, contrary to our hypothesis. The google-flan-t5-xl model yielded better results as the query model than gpt-3.5-turbo or LLaMA 2 70b-chat. However, the converse is not true; google-flan-t5-xl performed quite poorly as the knowledge base, proving that part of our hypothesis is correct. Overall, with a descriptive model generating the knowledge base and a general model querying it with pre-trained text, embedding models can adequately classify medical text data.

We performed this study to explore a pipeline of text embedding models in the medical space and we found it to be successful. However, we performed this study with only eight sufficiently different ailments, but there are many more ailments, often with a lot of overlap in symptoms. While we found that text embedding models are successful medical text classifiers on this small scale, further work is needed to determine whether this success transfers through scale. Ultimately, through our study, we found an effective way to classify medical text data while using existing models and without having to train multi-billion parameter LLMs. Additionally, we have introduced and validated a framework of text classification for the medical field that could enable a doctor to get diagnosis classifications for many ailments at once without having to interact conversationally with an LLM.

\bibliography{references}

\appendix
\section*{Appendix}
All 18 robustness tests can be viewed here: \href{https://github.com/bloodpool7/TextClassification}{Robustness Tests}

\end{document}